\documentclass[prb,twocolumn,showpacs]{revtex4}

\usepackage{graphicx}

\begin{document}

\title{Configurational entropy of network-forming materials}

\author{R.L.C. Vink}
\email{vink@phys.uu.nl}
\homepage{http://www.phys.uu.nl/~vink}
\affiliation{
        Institute for Theoretical Physics,
        Utrecht University,
        Leuvenlaan 4,
        3584 CE Utrecht, the Netherlands }

\author{G.T. Barkema}
\affiliation{
        Institute for Theoretical Physics,
        Utrecht University,
        Leuvenlaan 4,
        3584 CE Utrecht, the Netherlands }

\date{\today}

\begin{abstract}
\vspace{0.5cm} We present a computationally efficient method to calculate
the configurational entropy of network-forming materials. The method requires
only the atomic coordinates and bonds of a single well-relaxed
configuration. This is in contrast to the multiple simulations that are
required for other methods to determine entropy, such as thermodynamic
integration. We use our method to obtain the configurational entropy of
well-relaxed networks of amorphous silicon and vitreous silica. For these
materials we find configurational entropies of $1.02 k_B$ and $0.97 k_B$
per silicon atom, respectively. 
\end{abstract}


\maketitle


In materials such as vitreous silica, amorphous silicon or vitreous ice,
the structure is determined by the set of bonds (covalent or hydrogen)
between particles. In this manuscript, we will refer to these materials as
{\it network-forming materials}.  While the local environment of each
particle is usually governed by strict rules, the bonded network can
show a wide variety of different topologies. The focus of this work
is to present a method to estimate the number of topologies that a
network-forming material can take, or more precisely, its configurational
entropy.

The common computational procedure to estimate the entropy $S$ at
temperature $T_2$ is to measure the energy $E$ as a function of
temperature and then to integrate from a temperature $T_1$ at which the
entropy is known:
\begin{equation}\label{eq:c}
  S(T_2)-S(T_1)=\int_{T_1}^{T_2} \frac{1}{T} \frac{\partial E}{\partial T} dT.
\end{equation}
This requires sampling a large number of different networks and can
therefore only be applied to systems with fast dynamics.  In this work,
we present a completely different approach to estimate the entropy, based
on information theory~\cite{shannon}, and related to the work of Schlijper
{\it et al.} who determined the entropy of the Ising and three-states
Potts models~\cite{schlijper}.

One important concept in information theory is the Shannon entropy. It is
commonly explained in the context of a string of $n$ bits. In this case,
the Shannon entropy $H(n)$ is defined as:
\begin{equation}\label{eq:H}
  H(n) = -\sum_i p(i) \log_2 p(i),
\end{equation}
where the index $i$ runs over all possible bit sequences of length $n$ and
$p(i)$ is the probability of sequence $i$ occurring. A related concept is
the entropy density $s$ of a large string of $N$ bits:
\begin{equation}\label{eq:slim}
  s = \lim_{n \rightarrow \infty} \left[ H(n+1)-H(n) \right].
\end{equation}
A practical procedure to estimate $s$ for such a long string is to
extract from it a large number $m$ sub-sequences, each containing
$n$ bits with $n \ll N$. An estimate for the probabilities $p(i)$ is
then given by:
\begin{equation}\label{eq:count}
  p(i) \approx \frac{f_i}{m},
\end{equation}
where $f_i$ equals the number of times sub-sequence $i$ was observed. The
estimates for $p(i)$ in combination with Eq.~(\ref{eq:H}) yield $H(n)$.
The entropy density $s$ is then obtained using Eq.~(\ref{eq:slim}).
Usually, the limit $n \rightarrow \infty$ converges rapidly and even
moderate values of $n$ are sufficient to predict $s$ accurately.


For systems in equilibrium it is easily shown that the Shannon entropy and
the thermodynamic entropy of Eq.~(\ref{eq:c}) are equivalent, apart from a
factor of $\ln(2)$. In this case, the probabilities $p(i)$ are simply the
Boltzmann weights: $p(i) = \exp(-\beta E_i)/Z$, where $\beta$ is the
inverse temperature, $Z$ the partition function and $E_i$ the energy of
state $i$. In the present work we show how information theory can also be
used to obtain the configurational entropy of network forming materials.
The method we present can be applied to any network provided (1) the
atomic coordinates are known and (2) a list of bonds is supplied or can be
constructed --- for instance based on a distance criterion --- which
uniquely determines the network.

To determine the configurational entropy of a network-forming material
we choose a large number $m$ of random positions in the simulation
cell. For each position we find the nearest $n$ particles and identify the
graph formed by the bonds connecting these particles.  We then assign a label
to this graph, based on the graph automorphism~\cite{nauty}. We count
the number of times a graph is observed and use Eq.~(\ref{eq:count}) to
estimate its corresponding probability of occurrence. These probabilities
are fed into Eq.~(\ref{eq:H}) to obtain $H(n)$.

A small displacement of one of these random positions will usually result
in exactly the same list of $n$ nearest particles and thus the same graph.
Consequently, there is an upper bound to the number $m$ of random
positions that one should choose in a simulation cell containing $N$
particles. An estimate of this upper bound is obtained from the typical
distance over which a random position can be displaced without altering
the selected graph. From this we obtained as upper bounds $m=1.6 n N$ and
$m=3.4 n^2 N$ for two-dimensional and three-dimensional networks,
respectively.

A related side-effect of choosing random positions is that the number of
different graphs observed gets multiplied with a factor proportional to
$n^{d-1}$. This results in a correction to $H(n)$ of the form
$g(n)=(d-1)\log(n)$, with $d$ the spatial dimension of the network. The
latter can be verified in crystalline networks where the configurational
entropy is zero.

Graphs with a probability $p(i)$ smaller than $1/m$ will likely be
observed only once, if at all. This finite-size effect grows with $n$,
when the selected graphs become very complex. To monitor the impact of
this effect, we record the quantity $H_1(n)$, defined as the contribution
to $H(n)$ of the topologies observed once. We reject the measurements for
which $H_1(n)$ exceeds one percent of $H(n)$. Furthermore, we observed
that the quantity $H(n)+c \cdot H_1(n)/H(n)$ converges much faster with
increasing $m$ than $H(n)$ itself, with a suitable choice of $c$.  We
therefore use this extrapolated value as our best estimate for $H(n)$.

\begin{figure}
\begin{center}
\includegraphics[width=9cm]{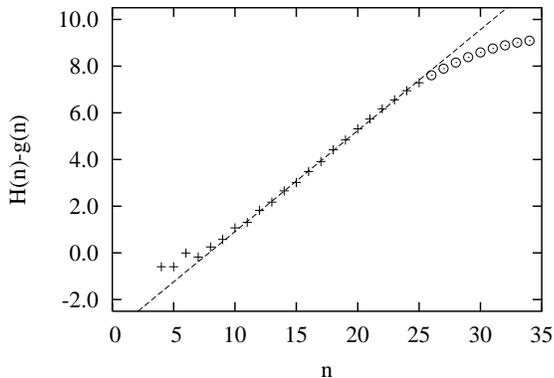}
\caption{\label{fig:fit} Typical behavior of the corrected Shannon entropy
$H_c(n)$ in units of $k_B$ as a function of graphsize $n$; crosses
(circles) mark measurements for which $H_1(n)$ contributes less (more)
than one percent to $H(n)$. The dashed line is a straight-line fit to the
crosses, starting from $n=10$. The slope of this line is our estimate for
the configurational entropy per atom $s$. These data are obtained from a
two-dimensional sillium configuration containing 20,000 atoms.}
\end{center}
\end{figure}

In Fig.~\ref{fig:fit} we show the typical behavior of the corrected
Shannon entropy $H_c(n) \equiv H(n)-g(n)$ as a function of graphsize
$n$. The entropy follows from Eq.~(\ref{eq:slim}). The rapid convergence
of the limit is demonstrated by the linear behavior of $H_c(n)$ for
intermediate $n$ shown in Fig.~\ref{fig:fit} by the dashed line. The
entropy per atom equals the slope of this line. In Fig.~\ref{fig:fit} we
have marked with circles the points for which $H_1(n)$ exceeds one percent
of $H(n)$. These points suffer from finite-size effects and should not be
used.

\begin{figure}
\begin{center}
\includegraphics[width=7cm]{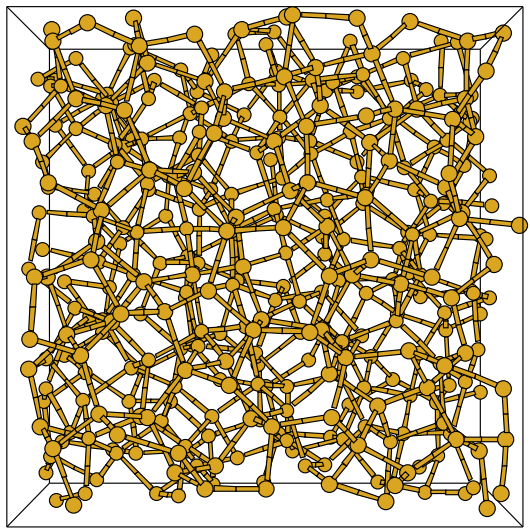}
\caption{\label{fig:3dconfig} A three-dimensional sillium network.
Each particle is four-fold coordinated but no long-range order exists.}
\end{center}
\end{figure}

\begin{figure}
\begin{center}
\includegraphics[width=7cm]{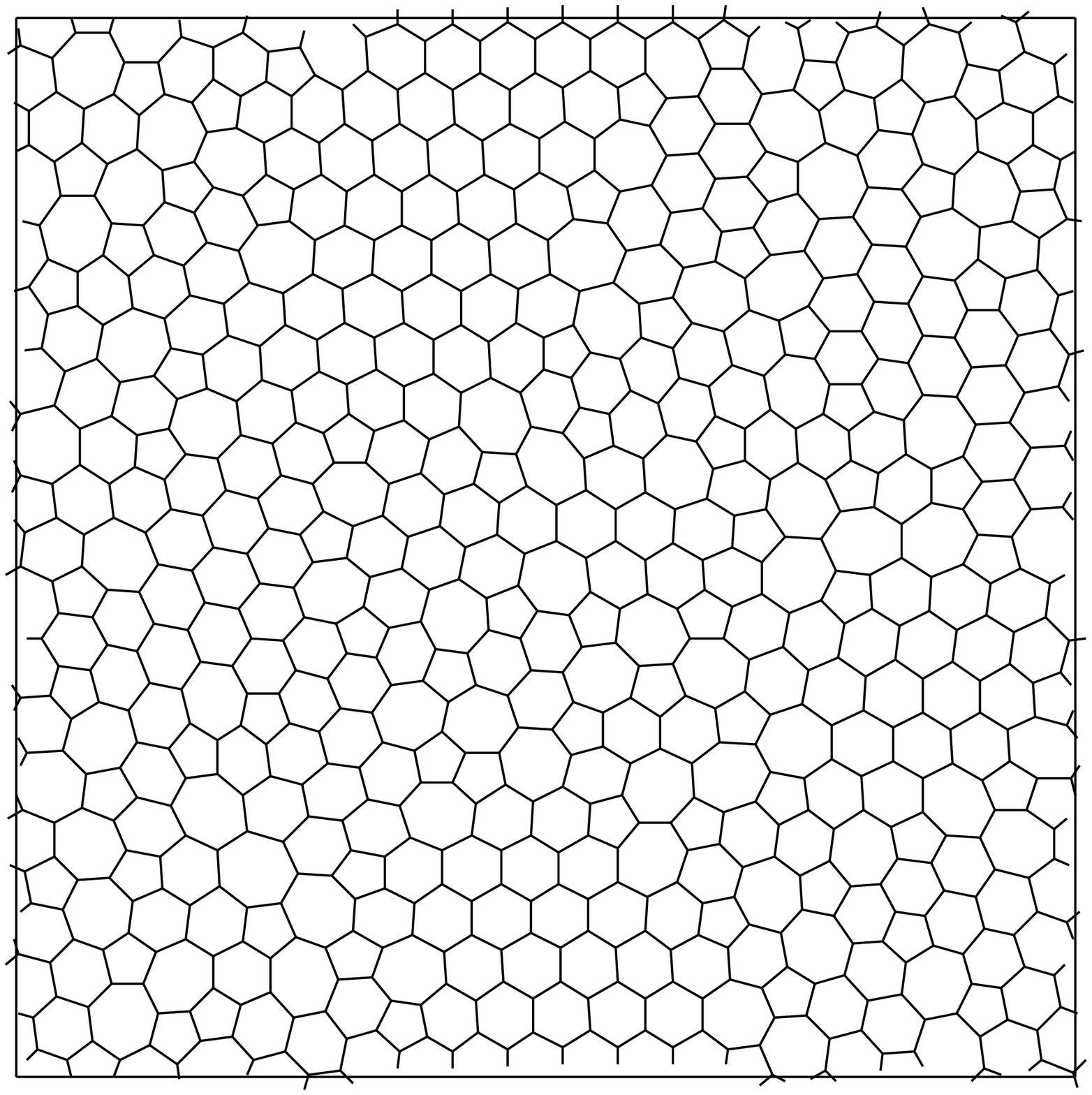}
\caption{\label{fig:2dconfig} A two-dimensional sillium network. Each
particle is three-fold coordinated, with a preferred bond angle of 120 degrees.}  
\end{center} 
\end{figure}


To test the usefulness of the above procedure, we apply it to the {\it
sillium} model~\cite{www}, one of the prototype models to study
network-forming materials.  In this model, tuned for amorphous silicon, an
explicit list of covalent bonds between pairs of Si atoms is kept, with
the property that each Si atom is bonded to four neighboring atoms. The
energy is described by the Keating potential, which contains a quadratic
penalty for bond-length deviations from the crystalline distance of 2.35
\AA, and a quadratic penalty for bond-angle deviations from the
tetrahedral angle $\Theta_0=\arccos(-1/3)$. The list of bonds determines
the atomic positions uniquely, since in this model the energy is minimized
at all times.  As a result, the phase space of this model is limited to a
finite number of $3N$-dimensional points. The evolution of the network
consists of a large number of random bond transpositions, each accepted
with the Metropolis probability:
\begin{equation}
  P = \min \left[1, \exp \left( 
	\frac{E_b-E_f}{k_B T} \right)  \right],
\end{equation}
where $T$ is the temperature and $E_b$ and $E_f$ are the total (minimized)
energies of the system before and after the proposed bond transposition. A
typical sillium network is shown in Fig.~\ref{fig:3dconfig}. Our networks
are generated as described in Refs.~\onlinecite{barkemahq,vinkhq}, in
which a number of algorithmic improvements have been proposed as compared
to the original algorithm of Wooten, Winer and Weaire.

To verify the validity of the information theory approach we first turn to
a two-dimensional version of the sillium model. Contrary to the
three-dimensional sillium model, the dynamics here does not lead to glassy
states.  This allows us to determine the entropy in equilibrium using both
the standard approach --- via Eq.~(\ref{eq:c}) --- as well as information
theory. In this two-dimensional model, atoms are three-fold coordinated
and the ideal bond angle is 120 degrees; a typical configuration is shown
in Fig.~\ref{fig:2dconfig}.

We simulate a two-dimensional sillium network containing 1008 atoms for a
number of (physically interesting) temperatures ranging from $k_B T=0.15$
eV to $k_B T=0.60$ eV using periodic boundary conditions in both
directions.  At each temperature we bring the system to equilibrium with
100 attempted bond-transpositions per atom. Next, 20 snapshots of the
network are stored; each separated by five attempted bond transpositions
per atom. From these snapshots the average energy and the Shannon entropy
are obtained. In determining the Shannon entropy we draw clusters from
each snapshot simultaneously. After the simulation has covered the entire
temperature range, the entropy is also obtained from the average energy
measurements and Eq.~(\ref{eq:c}). 

\begin{figure} 
\begin{center} 
\includegraphics[width=9cm]{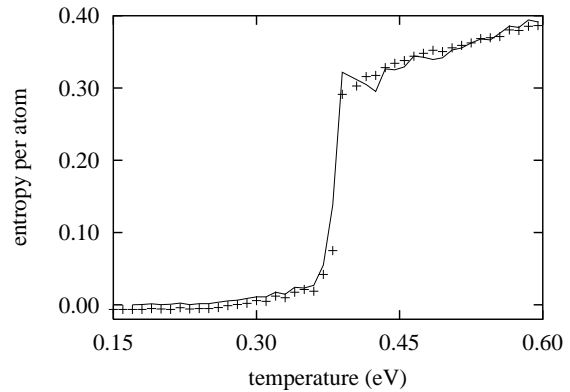}
\caption{\label{fig:comp} Diagram showing the configurational entropy per
atom in units of $k_B$ as a function of temperature in eV for a
two-dimensional sillium network consisting of 1008 atoms. The solid curve
shows the entropy per atom as obtained using the standard thermodynamic
approach; crosses show the entropy per atom obtained using information
theory.}
\end{center}
\end{figure}

In Fig.~\ref{fig:comp} we show the configurational entropy as a function
of temperature. The solid curve shows the entropy per atom as obtained
using the standard thermodynamic approach; crosses show the entropy per
atom as obtained using information theory. There appears to be a phase
transition near $k_B T=0.37$ eV shown by the sudden jump in entropy. The
remarkable feature of the displayed results is the overall good agreement
between the two methods, even near the phase transition.


Next, we use information theory to determine the configurational
entropy of a well-relaxed three-dimensional sillium network containing
20,000 atoms.  This network was generated using the improved Wooten,
Winer and Weaire algorithm~\cite{barkemahq,vinkhq}. The Keating energy of
this network is 0.286 eV per atom; the standard deviation in the mean
bond-angle is 9.63 degrees. Structural and electronic properties of
this network are in excellent agreement with experimental properties
of amorphous silicon~\cite{vinkhq}.  Applying our method to this model,
we obtain a configurational entropy of $1.02 k_B$ per atom.

The (classical) vibrational entropy is obtained from the eigenvalues of the
dynamical matrix and for this network was found to be $3.07 k_B$ per atom.
For the crystalline phase the configurational entropy is zero and only the
vibrational entropy contributes to the entropy, in this case $3.18 k_B$
per atom. The difference in entropy between the crystalline and amorphous
phases of silicon is therefore estimated to be $0.91 k_B$ per atom.

The energy difference between the crystalline and amorphous phases
of silicon was recently determined by Biswas~\cite{biswas} using
tight-binding, who found an energy difference of 0.18 eV per atom.

With the above estimates for the differences in energy and entropy we can
calculate the transition temperature $T_c$, defined as the temperature
where the free energy difference between the crystalline phase and the
amorphous phase changes sign. It is given by:
\begin{equation}\label{eq:tc}
  T_c = \frac{\Delta E}{\Delta S},
\end{equation} 
where $\Delta E$ and $\Delta S$ are the energy difference and the
entropy difference between the crystalline and the amorphous phase,
respectively.  Substitution of our estimates for $\Delta E$ and $\Delta S$
into Eq.~(\ref{eq:tc}) yields \mbox{$T_c\approx 2300$K}. This temperature
compares well with the value of \mbox{$T_c \approx 2500$K} as inferred
from calorimetric experiments~\cite{sinke}, and is well above the melting
point of silicon. It confirms that the amorphous phase is not
thermodynamically stable at any temperature.


As a final application of our method we determine the configurational
entropy of vitreous silica. The structure of this material is formed by
covalent bonds between silicon atoms and oxygen atoms. Barring rare
defects, each silicon atom is bonded to four oxygen atoms and each oxygen
atom is bonded to two silicon atoms. We generate a silica network
containing 3000 atoms (with periodic boundary conditions) in a spirit
similar to amorphous silicon, but we replace the Keating potential by the
one of Tu and Tersoff~\cite{tu1,tu2}. The resulting network is then
quenched with the more realistic BKS potential~\cite{beest}, using
parameters as described in Ref.~\onlinecite{vollmayr}. The BKS energy per
silicon atom of the quenched network is \mbox{0.13 eV} above that of the
$\beta$-cristobalite structure. For comparison, samples prepared by
molecular dynamics typically yield a much larger energy difference of
\mbox{0.30 eV}. This clearly demonstrates that our network is well
relaxed. After quenching with the BKS potential, the atoms remain
perfectly coordinated: the silicon-oxygen bonds are easily reconstructed
from a distance criterion and the silicon-oxygen radial distribution
function. For our network, the average \mbox{O-Si-O} bond-angle is
109.43 degrees with a standard deviation of 4.40 degrees; the average
\mbox{Si-O-Si} bond-angle was found to be 150.85 degrees with a standard
deviation of 12.09 degrees.

To estimate the configurational entropy per silicon atom in well-relaxed
vitreous silica, we used the above network and replaced each oxygen atom
plus its two bonds by a single silicon-silicon bond. Next, we applied the
information-theoretic method described earlier on this network and found
for the configurational entropy $0.97 k_B$ per silicon atom.  Given the
limited size of this sample we estimate that the actual entropy might
be up to 10\% larger. Ignoring the difference in vibrational entropy, as
well as the entropy contribution of possible multiple oxygen positions,
we obtain for the transition temperature of vitreous silica \mbox{$T_c
\approx 1600$K}.

In summary, we have developed a computationally efficient method to
determine the configurational entropy of network forming materials.  For
well-relaxed samples of amorphous silicon and vitreous silica, we find
for the entropy per silicon atom $1.02 k_B$ and $0.97 k_B$, respectively.
In future research, we hope to extend the applicability of this method to
other disordered materials such as colloidal systems and metallic glasses.


We thank Henk van Beijeren and Normand Mousseau for stimulating discussion,
and Normand Mousseau also for the computation of the BKS energies.

\bibliographystyle{prsty}

\begin{thebibliography}{99}

\bibitem{shannon} C.E. Shannon, The Bell System Technical Journal {\bf
27}, 379-423, 623-656 (1948).

\bibitem{schlijper} A.G. Schlijper, A.R.D. van Bergen, and B. Smit, Phys.
Rev. A {\bf 41}, 1175 (1990).

\bibitem{nauty} B.D. McKay, Nauty User's Guide (Version 1.5), Technical
Report TR-CS-90-02, Australian National University, Department of Computer
Science, 1990.

\bibitem{www} F. Wooten, K. Winer, and D. Weaire, Phys. Rev. Lett. {\bf
54}, 1392 (1985).

\bibitem{barkemahq} G.T. Barkema and N. Mousseau, Phys. Rev. B {\bf 62},
4985 (2000).

\bibitem{vinkhq} R.L.C. Vink, G.T. Barkema, M.A. Stijnman, and R.H.  
Bisseling, Phys. Rev. B {\bf 64}, 245214 (2001).

\bibitem{biswas} P. Biswas, private communication.

\bibitem{sinke} W.C. Sinke, S. Roorda, and F.W. Saris, J. Mater. Res.  
{\bf 3}, 1201 (1988).

\bibitem{tu1} Yuhai Tu, J. Tersoff, G. Grinstein, and D. Vanderbilt,
Phys. Rev. Lett. {\bf 81}, 4899 (1998).

\bibitem{tu2} Yuhai Tu and J. Tersoff, Phys. Rev. Lett. {\bf 84}, 4393
(2000).

\bibitem{beest} B.W.H. van Beest, G.J. Kramer, and R. A. van Santen, Phys.  
Rev. Lett. {\bf 64}, 1955 (1990).

\bibitem{vollmayr} K. Vollmayr, W. Kob, and K. Binder, Phys. Rev. B {\bf
54}, 15808 (1996).

\end{thebibliography}

\end{document}